\documentstyle[aps,multicol,psfig,epsf,epsfig]{revtex}            
\begin{document}
\def\pa{\parallel}
\def\pe{\bot}
\let\a=\alpha \let\b=\beta  \let\c=\chi \let\d=\delta  \let\e=\varepsilon
\let\f=\varphi \let\g=\gamma \let\h=\eta \let\k=\kappa  \let\l=\lambda
\let\m=\mu   \let\n=\nu   \let\o=\omega    \let\p=\pi
\let\r=\varrho  \let\s=\sigma \let\t=\tau   \let\th=\vartheta
\let\y=\upsilon \let\x=\xi \let\z=\zeta
\let\D=\Delta \let\F=\Phi  \let\G=\Gamma  \let\L=\Lambda \let\Th=\Theta
\let\O=\Omega       
\newcommand{\ie}{\hbox{\it i.e.\ }}         
\draft
\tightenlines
\title{Is the particle current a relevant feature in
 driven lattice gases?}
\author{A. Achahbar $^1$, 
Pedro L. Garrido $^2$, J. Marro $^2$, and Miguel A. Mu\~noz$^{2}$}
\address{
$^1$ Departement de Physique, Faculte des Sciences, B.P. 2121
M'hannech,  93002 Tetouan, Morocco. \\
$^2$ Institute {\em Carlos I} for Theoretical and Computational Physics\\
and Departamento de Electromagnetismo y F\'\i sica de la Materia,
Universidad de Granada,
 18071 Granada, Spain.\\
}
\date{\today}

\maketitle
\begin{abstract}
By performing extensive MonteCarlo simulations we show that
the infinitely fast driven lattice gas (IDLG)  
shares its critical properties with the randomly driven lattice gas (RDLG).
 All the measured exponents, scaling functions and 
amplitudes are the same in both cases. 
This strongly supports the idea that the main relevant non-equilibrium 
effect in driven lattice gases is the anisotropy 
(present in both IDLG and RDLG) 
and not the particle current (present only in the IDLG). 
This result, at odds with the predictions from the standard 
theory for the IDLG, supports a recently proposed alternative theory. 
The case of finite driving fields is also briefly discussed. 

\end{abstract}


\begin{multicols}{2}
\narrowtext
 The Ising model exhibits a prototypical equilibrium phase transition, 
and the associated $\phi^4$ Ginzburg-Landau theory is 
a paradigm of continuous theory for equilibrium critical 
phenomena \cite{HH,AZJ}. However, thermodynamic equilibrium is 
exceptional in nature where stationary states are typically
away from equilibrium \cite{MD}.
With the purpose of defining simple lattice models describing 
generic {\it non-equilibrium} phase transitions, different tentatives 
have been made in the last two decades. 
Among them, perhaps the most intriguing example is the 
driven lattice gas (DLG) \cite{MD,katz,zia}.
 (Other interesting examples are the directed percolation model 
\cite{MD} and the Kardar-Parisi-Zhang equation).
%\cite{KPZ}.
The DLG, being a straightforward extension of the Ising model, 
has in fact become a workbench for emergent 
non-equilibrium theories and field theoretical approaches. 

The DLG is a d-dimensional kinetic Ising model with conserved dynamics, 
in which transitions in the direction (against the direction) 
of an externally applied field, $\vec{E}$, are favored
(unfavored) \cite{MD,katz,zia}, while transitions perpendicular
to the field are unaffected by it.
The field induces two main non-equilibrium effects:  
(i) the presence of a net current of particles along its direction,
and (ii) {\it strongly} anisotropic configurations \cite{weak}.
At high temperatures, the system is in a disordered phase while,
for half-filled lattices (the only case we refer to in what
follows) there is a second-order critical point, 
below which the DLG segregates into (two) 
high and low density aligned-with-the-field stripes.
Establishing unambiguously the DLG universality class 
is an important issue in the way to rationalize the 
behavior of non-equilibrium systems. 

Continuous approaches such as Langevin and associated 
field theories \cite{AZJ} have been most useful
in studying universality issues in equilibrium critical phenomena. 
In particular, coarse-grained approaches
combined with renormalization group (RG) techniques 
provide a method for the classification of the different possible terms (operators)
 as relevant, irrelevant or marginal. 
In fact, Langevin equations are more illuminating than other 
(even more rigorous) approaches, as they permit to understand
systematically how possible perturbations or model 
variations would affect critical properties.
 Consequently, many studies have focused on the DLG and 
its universality by using both non-equilibrium continuous 
approaches and computer simulations
(unfortunately, general exact solutions are not available). 
Within this perspective, it is somewhat deceptive that after
many computer and analytical studies, the universality class
of the DLG remains a debated issue \cite{MD,zia,Rapid,micidiale}.

A phenomenological Langevin equation intended to capture 
the relevant physics of the DLG at criticality was proposed and 
renormalized more than a decade ago \cite{JS}. 
This equation, referred to as {\it driven diffusive system} 
(DDS), is a natural extension of 
the conserved $\phi^4$ theory for the Ising equilibrium transition
(model B \cite{HH}) and seems to capture the main symmetries and 
conservation laws of the discrete DLG. 
It includes a particle current term (which from naive power counting
turns out to be the most relevant nonlinearity) 
as well as anisotropic coefficients.
It certainly is a suitable and very reasonable candidate to be 
{\it the canonical coarse-grained model representative of the DLG universality class}. 
The DDS Langevin equation reads:
\begin{eqnarray}
\label{dds}
\partial_t\phi({\bf r},t)= &&
\tau_\pe \nabla_{\pe}^2\phi
-\nabla_{\pe}^4\phi
+{ \lambda \over6}\nabla^2_{\pe}\phi^3
 \nonumber \\
&& + \tau_\pa  \nabla_{\pa}^2 \phi
- \alpha  \nabla_{\pa} \phi^2
+ {\bf \z}({\bf r},t),
\end{eqnarray}    
where $\phi$ is the coarse grained field, ${\bf \z}$ is a conserved
Gaussian noise and the cubic term, being a dangerously irrelevant 
variable \cite{AZJ}, is kept in order to ensure stability \cite{JS}.
$\tau_{\pa}$, $\tau_\pe$, $\lambda$ and $\alpha$ are model parameters.
The most emblematic (exact) prediction derived 
from the DDS RG-analysis, namely, the mean field behavior of the 
order parameter critical exponent, $\beta=1/2$, \cite{JS}
has eluded a large number of Monte Carlo (MC) analysis 
aimed at probing it \cite{aclarar}, however.
In particular, systematic deviations from the predicted 
scaling are observed both in $d=2$ \cite{MD,FSS} 
and in $d=3$ \cite{LW,aclarar}.
Indeed, different MC analysis (performed using
a variety of aspect-ratios and order parameters) lead systematically 
to a value of $\beta$ close to $\approx 0.3$ (in $d=2$),
with error bars excluding apparently the value $\beta=1/2$
(see \cite{MD} for a critical review of simulation analysis).
This is a main indication that, strikingly enough 
{\it the DDS equation does not describe properly the infinitely fast 
driven DLG (IDLG) critical properties}.
Moreover, there are some other hints suggesting strongly that
the differences between the predictions of the standard Langevin approach 
and MC results are more fundamental than a simple discrepancy in the 
value of $\beta$.
In particular, the intuition developed from MC simulations 
of the DLG and variants of it \cite{MD} suggests that, 
contrarily to what the DDS equation establishes,
{\it it is the anisotropy and not the presence of a current the  
basic ingredient controlling the critical behavior} \cite{weak}.
For instance, in a modified DLG in which anisotropy is included   
by means others than a current \cite{ALGA}, the scaling behavior at 
criticality remains unaltered upon switching 
on an (infinite) driving (see \cite{ALGA,MD}).
Other compelling evidences supporting this hypothesis can be found
in \cite{MD,MAGA}.     

In an attempt to clarify this puzzling situation, and reconcile continuous
approaches with numerics, different scenarios have been explored.
In particular, an alternative route to  build up Langevin equations 
starting from generic microscopic master equations was recently 
proposed \cite{Rapid}.
By applying this approach to the DLG, one observes that,
owing to a transition-rates saturation
effect, the coefficient $\alpha$ 
of the non-linear current term, $\nabla_{\pa} \phi^2$,
vanishes in the limit of infinite driving fields and, therefore, 
it does not appear in the final Langevin equation nor it is generated 
perturbatively \cite{Rapid}.
The resulting  theory (alternative to Eq.(\ref{dds})) is:
%\begin{eqnarray}
\begin{equation}
\label{ads}
\partial_t\phi({\bf r},t)= 
\tau_\pe \nabla_{\pe}^2\phi
-\nabla_{\pe}^4\phi
+{\lambda \over6}\nabla^2_{\pe}\phi^3
% \nonumber \\
 + \tau_\pa \nabla_{\pa}^2 \phi
+ {\bf \z}
\end{equation}     
plus higher order irrelevant contributions (note that a linear current term
has been eliminated by employing a Galilean transformation \cite{zia,Rapid}).
This equation, named below {\it anisotropic diffusive system} (ADS), 
is a well known one: it coincides with the Langevin 
equation representing the random DLG 
(RDLG) \cite{RDDS,RDDS2} (for which
the driving field takes values $\infty$ and $-\infty$ 
in a random unbiased fashion, generating anisotropy 
but not an overall current). 
This theory has been extensively studied 
in \cite{RDDS,RDDS2}; its critical dimension is $d_c=3$ 
(instead $d_c=5$ for the DDS) and the critical exponents 
and finite size scaling (FSS) properties are now well known.
Other systems in this universality class are the two-temperature model
\cite{2T} and the ALGA model \cite{ALGA}.
This theory for the IDLG includes anisotropy as its basic 
non-equilibrium ingredient. Instead - for non-saturating, 
finite, driving fields - the cancellation of the nonlinear current term does 
not occur, and our method recovers the standard DDS equation.

Aiming at further clarifying these issues, 
we report here on extensive MC simulations of the 
IDLG and the RDLG in $d=2$. 
The main objectives are:  
(i) trying to conclude whether the IDLG and the RDLG 
share the same critical behavior or not; and
(ii) measuring the critical exponents by performing systematic 
anisotropic finite-size scaling (FSS).
In fact, we perform FSS analysis for both the IDLG and the RDLG 
 by following the anisotropic 
FSS scheme proposed in \cite{RDDS} consistent with
the ADS theory;
this allows us to analyze systematically possible
scaling differences between both models. We also report on
the case of finite-driving DLG.

%\vspace*{-0.5cm}
We consider rectangular lattices of size 
$L_\pa \times L_\pe$ with periodic boundary conditions and
random sequential updating \cite{MD,zia};
 the external field $\vec{E}$ acts in the $x$ (parallel) direction. 
Particles jump to a randomly chosen nearest neighbor site 
(provided that it is empty) 
with probability:
$\min(1, \exp[-\beta ( \Delta H + E \Delta j)])$, where $\Delta H$ is 
the energy (Ising Hamiltonian) variation, and $\Delta j= (-1, 0, 1)$ for 
jumps along, against, and orthogonal to the direction of the field,
respectively. 
Following \cite{FSS,RDDS2} the order parameter is chosen as the 
structure factor $S(0,2\pi/L_\pe)$.
%\begin{equation}
%m={1 \over 2 ~ L_\pa} \sin({\pi \over L_\pe}) \left|
%\sum_{x=1}^{L_\pa}   \sum_{y=1}^{L_\pe} \sigma_{x,y} e^2 \pi i
%/L_\pe \right|
%\label{mag}
%\end{equation}
%where $\sigma_{x,y} = 0$ or $1$ are the occupation variables). 
In order to perform a systematic anisotropic FSS we considered
system sizes 
$20 \times 20$, $45 \times 30 $, $80 \times 40$, and $125 \times 50$.
These aspect ratios satisfy $ L_\pa^{\nu_\pe/\nu_\pa} 
= 0.2236 \times L_\pe $, where $\nu_\pe/\nu_\pa \approx 1/2$ consistent
with ADS anisotropic spatial scaling \cite{FSS,RDDS2}.
The number of MC steps considered varied between 
$1.8 \times 10^8$ and $2.4 \times 10^8$, much
larger than in any previously reported MC simulations.
The total CPU time employed is about eight months in a pentiumIII
400Mhz machine.
The critical temperature is determined by using the 
fourth (Binder) cumulant method \cite{Binder}.
For the IDLG, the critical temperature is found to be  
$T_c^{I}=1.396(4) T_o$ ($T_o$ is the Onsager temperature),
slightly below previously reported values \cite{zia,MD}, 
while we find $T_c^{R}=1.390(4) T_o$ for the RDLG 
(see insets of Fig.3). 
These critical values were employed for the FSS analysis. 
 In Fig.1 we plot the order parameter, rescaled by a factor 
$L_\pa^{\beta/\nu_\pa}$, versus $\epsilon L_\pa^{1/\nu_\pa}$, where
$\epsilon$ is the distance to the critical point, for different system 
sizes $L_\pa$. A nearly perfect data-collapse is obtained by fixing
 $\nu_{\pa} = 1.25$ and $\beta=0.33$. The collapse gets worse upon
slightly changing these values; more precise estimates of the associated 
error-bars is a difficult and not essential issue in this context.
Notice that we are plotting in the same
graph data for the IDLG and for the RDLG, implying that the FSS 
scaling function is precisely the same for both models. 
Furthermore, the slopes 
of the asymptotic branches are approximately $1/3$ and $-0.61$, 
consistent both with the order parameter exponent being $\beta \approx 0.33$, and
$\gamma \approx 1.22$ (see below) \cite{sf}.  
In general, even when the scaling functions are universal their 
corresponding amplitudes are not expected to be so. For this reason, 
usually one has to introduce the, so called, {\it metric factors} 
(varying amplitudes) \cite{metric} in order to obtain superposition 
of scaling functions within the same universality class.
 Contrarily to this expectation, the magnetization scaling functions for 
IDLG and RDLG overlap perfectly. Therefore, it comes as a surprise
that not only the scaling functions and the $\beta$ exponent are
universal in both models, but even the amplitudes coincide. A similar situation
has been recently reported for a different type of anisotropic FSS \cite{H}.
\vspace{-0.75cm}
\begin{figure}
\centerline{
\epsfxsize=8.5cm
\epsfysize=6cm
\epsfbox{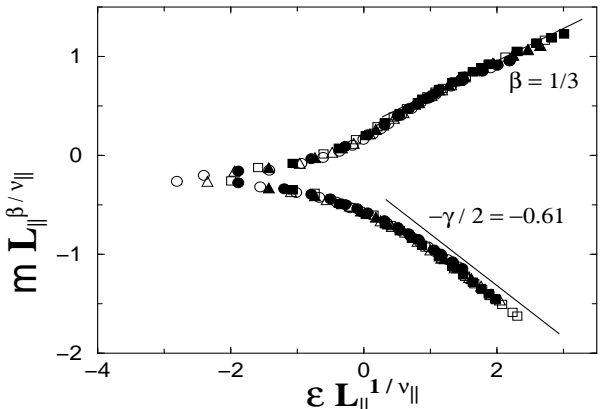}}
\caption{Log-log plot of the order parameter rescaled by
$L_{\pa}^{\beta/\nu_\pa}$ vs. $\epsilon L_{\pa}^{1/\nu_\pa}$,
for different system sizes: $\bigcirc 45 \times 30$,
$\bigtriangleup 80 \times 40$, $\Box 125 \times 50$.
Filled (empty) symbols stand for the RDLG (IDLG); errorbars are
smaller than the symbols.}
\label{Figure1}
\end{figure} 
\vspace{-1cm}
\begin{figure}
\centerline{
\epsfxsize=8.5cm
\epsfysize=6cm
\epsfbox{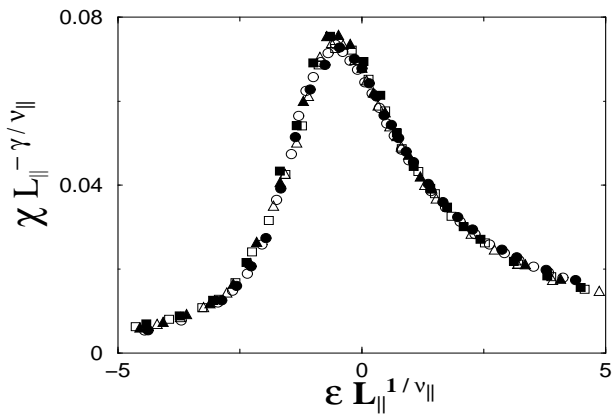}}
\caption{Log-log plot of the susceptibility rescaled by
 $L_{\pa}^{-\gamma/\nu_\pa}$ vs. $\epsilon L_{\pa}^{1/\nu_\pa}$.
Symbols are as in Fig.1 (larger than errorbars).}
\label{Figura2}
\end{figure} 
\vspace{-0.25cm}
We have also computed the system susceptibilities,
defined as the relative fluctuations of the order parameter:
% around its average value:
$\chi = {L_\pa \over sin(\pi/L_\pe)} [ < m^2 > - < m >^2 ]$.
In Fig.2, we plot the susceptibility times
$L_{\pa}^{-\gamma/\nu_\pa}$ as a function of the rescaled distance to 
the critical point, $\epsilon L_{\pa}^{1/\nu_\pa}$. The best
data collapse is obtained by employing the values $\gamma=1.22$ and
$\nu_\pa=1.25$ for both models with, again, coinciding amplitudes.
It should be stressed that this is the first time a really good collapse
is observed below the critical point for anisotropic scaling of
 the IDLG.
 Plotting the dimensionless Binder cumulant as a function of 
the rescaled distance to the critical point with $\nu_\pa=1.25$, 
again, nearly perfect data collapse is obtained for 
both models and all system sizes (Fig. 3).
We also performed simulations in square lattices 
($128 \times 128$) as in some previous studies \cite{MD,MAGA}. 
Monitoring $m^{1/\beta}$ as a function of $T/T_o$ we see no appreciable 
systematic difference between the curves for IDLG and RDLG, 
that have the same slope within numerical accuracy.
The best linear fit correlation is obtained for $\beta \approx 0.33$ 
in both cases, providing an extra consistency check for our results.
Moreover, eye inspection of IDLG and RDLG configurations, 
for any geometry, at a fixed relative temperature, 
does not permit to distinguish one from the other. In particular, the
interfacial properties look identical. 
 Let us also stress that all the obtained exponent values are compatible 
with previous measures for the RDLG, 
as well as with the exponents obtained within
an $\epsilon-$expansion of the ADS theory \cite{RDDS,RDDS2}. 
\vspace{-3.25cm}
\begin{figure}
\centerline{
\epsfxsize=11cm
\epsfysize=16cm
\epsfbox{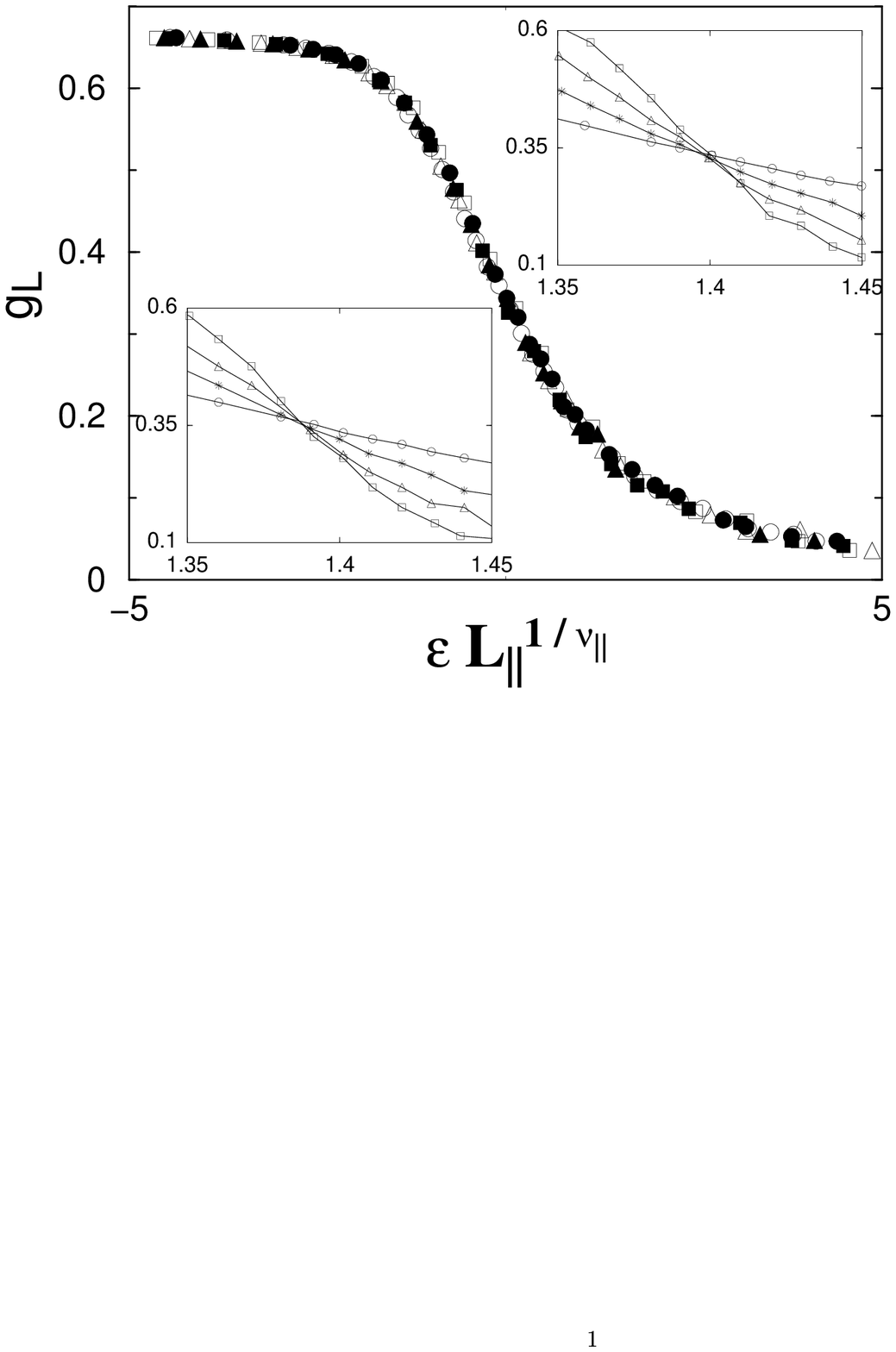}}
\vspace{-7.25cm}
\caption{Scaling plot of the fourth cumulant vs.
 $\epsilon L_{\pa}^{1/\nu_\pa}$. Upper (lower) inset: 
fourth cumulant for the IDLG (RDLG) vs. $T/T_o$. Symbols are as in Fig.1.}
\label{Figure3}
\end{figure}
\vspace{-0.25cm}
In conclusion, {\em MC results support strongly that both the 
IDLG and the RDLG belong in the same universality class,
and share not only critical exponents and scaling functions, 
but also the scaling amplitudes}.
This universality class is described by the ADS equation, Eq.(\ref{ads}). 
There is absolutely no hint of any difference in the asymptotic behavior
between the model with a current (IDLG) with respect to the 
current-less one (RDLG). 
{\it All the numerical evidence confirms that it is the anisotropy 
and not the net current (if any) the most relevant non-equilibrium
 ingredient of driven systems}.
As discussed in the introduction, this is striking from a
field theoretical perspective given that the nonlinear current term, 
$\nabla_\pa \phi^2$, 
is naively a relevant perturbation at the ADS fixed point.
In an alternative approach, the coefficient of such a term vanishes. 
In this picture, the fast drive limit corresponds to a sort 
of {\it multicritical point} in which an {\it a priori} relevant operator 
is absent due to a cancellation of its coefficient and, consequently 
the usual "up-down" Ising symmetry 
(i.e. the three-point correlation functions vanish)
is restored at criticality.  
In any case, it should be stressed that,
from a more general perspective, field theoretical descriptions 
of non-equilibrium systems are much more delicate and subtle 
than their equilibrium counterparts, and an extremely
careful inspection of the system symmetries, conservation laws, and dynamical
features is required before venturing to make predictions
on nonequilibrium universality issues. 
For example, fermionic and bosonic non-equilibrium
systems with the same dynamics, symmetries and conservation laws have recently
reported to belong to different universality classes \cite{Korea}. 
One could wonder whether 
the DLG hard-core interaction should 
be taken into account into its Langevin representation.
 
 Elucidating the critical behavior for finite $E$
remains a challenging and interesting objective.
Both, our alternative Langevin-building approach and the standard
one, Eq.(\ref{dds}), include a relevant current term in this case
and, consequently, predict $\beta=1/2$.
Obtaining clear-cut results in this case
is a computationally expensive task since:
(i) As the external field appears in the argument of an exponential,
even relatively small values of $E$ generate situations close 
to saturation, and strong crossover effects could hide the true 
asymptotic regime.
(ii) If the field value is taken too small, crossovers from the
equilibrium regime may also burden observations.           
A possible scenario that could follow from MC analysis is that
finite fields show mean field behavior; 
that would be a strong backing for our theory 
\cite{Rapid} that predicts the finite and
the infinite driving cases to be qualitatively different.
If, instead, scaling happens to be that of the ADS (as our preliminary
MC results for $E=3$ and $E=0.5$ seem to indicate; 
for $E=0.25$ results do not quite fit this indication)                
it would prove that it is for any arbitrary value of the driving field 
that anisotropy is the most relevant ingredient of driven systems. 
This scenario would uncover a new puzzling situation 
and would certainly call for deeper theoretical understanding. 
Huge and careful simulations 
would be required to extract neat conclusions 
overcoming difficulties (i) and (ii) above.

 Summing up, we have performed extensive MC simulations of the IDLG
and the RDLG. By using anisotropic finite size
scaling techniques we have shown that both models belong to the 
same universality class: their critical exponents, scaling functions and
amplitudes are undistinguishable and coincide with those of the ADS equation.
 This result supports the conclusion that it is the presence 
of anisotropic coefficients, and not the particle current the most 
relevant ingredient in these non-equilibrium driven problems (at least in the
fast drive limit). Further theoretical efforts are certainly required 
in order to (i) to sort out if our alternative Langevin approach is 
correct and what are its possible limitations, and (ii) 
to further clarify the universality issues of this
quintessential non-equilibrium problem.
Finally, it would also be very interesting to combine the powerful 
finite size methods recently introduced in this context by Caracciolo et al.
in a nice recent work \cite{pisa} with our alternative theory to 
verify if they lead to better data collapse than when used
to test the standard DDS equation (hopefully without 
having to introduce strong corrections to scaling and providing 
good order-parameter scaling).  
\vspace{0.15cm}

{\it ACKNOWLEDGMENTS-}
We are grateful to 
S. Caracciolo for his very valuable comments. We also 
thank F. de los Santos and Prof. A. Sekkaki.
This work has been supported by the European Network 
Contract ERBFMRXCT980183 and by the MCYT
%Ministerio de Educaci\'on
under project 
%DGESEIC,
 PB97-0842.

\newpage
\end{multicols}

\begin{thebibliography}{99}
\vspace*{-2.0cm}

\bibitem{HH}  P. C. Hohenberg and B. I. Halperin, Rev. Mod. Phys.
{\bf 49}, 435 (1977).

\bibitem{AZJ} 
%D. J. Amit, {\em Field Theory, the Renormalization 
%Group and Critical Phenomena},
%World Scientific, (Singapore, 1992).
 J. Zinn- Justin, {\it Quantum field theory
and critical phenomena}, Clarendon Press, (Oxford, 1994).     


\bibitem{MD}
J. Marro and R. Dickman,  {\it Nonequilibrium Phase Transitions
in Lattice Models
}, Cambridge University Press, (Cambridge, U.K., 1999).

\bibitem{katz}
S. Katz, J.L. Lebowitz and H. Spohn,
{\em Phys. Rev.\/} B {\bf 28},
 1655 (1983); {\em J. Stat. Phys.\/} {\bf 34}, 497 (1984).

\bibitem{zia}
B. Schmittmann and R. K. P. Zia, {\em Statistical Mechanics of
Driven Diffusive Systems\/}, in Phase Transitions and Critical
Phenomena, edited by C. Domb and J. Lebowitz (Academic, London, 1995)

%\bibitem{KPZ} T. Halpin-Healy and Y.-C. Zhang,
%Phys. Rep. {\bf 254}, 215 (1995).

\bibitem{weak} Anisotropy can also appear in equilibrium systems 
such as the Ising model with anisotropic couplings; 
R. Baxter, {\it Exactly solved models in Statistical Mechanics}, Academic Press, 
(London, 1982).
The term {\it weak anisotropy}, as opposed to {\it strong anisotropy},
applies to these cases, for which, 
owing to the fluctuation-dissipation relation, 
the equilibrium critical properties remain unaltered
\cite{zia}.           
           
\bibitem{Rapid} P. L. Garrido, M. A. Mu\~noz, and F. de los Santos,
Phys. Rev. E {\bf 61}, R4683 (2000).

\bibitem{micidiale} B. Schmittmann, et al. 
Phys. Rev. {\bf 61}, 5977 (2000).
          
\bibitem{JS}
H. K. Janssen and B. Schmittmann, {\em Z. Phys.\/} B {\bf 64}. 503
(1986). K.-t. Leung and J. L. Cardy, J. Stat. Phys. {\bf 44}, 567
(1986); ibid {\bf 45}, 1087 (Erratum) (1986).

\bibitem{FSS} K.-t. Leung, Phys. Rev. Lett. {\bf 66}, 453 (1991).
J. S. Wang, J. Stat. Phys. {\bf 82}, 1409 (1996).

\bibitem{LW}  K.-t. Leung et al.
% and J. S. Wang, 
Int. J. Mod. Phys. C {\bf 10}, 853 (1999).
               
\bibitem{aclarar} Other predictions of the standard
theory seem to be reasonably well verified by performing an adequate
finite size scaling analysis including a dangerously irrelevant scaling field
(see \cite{FSS} and \cite{LW}), 
but a fully consistent scaling, including good data collapse for 
the order parameter has never been observed. See the criticisms 
to \cite{FSS} and \cite{LW} in \cite{MD,ALGA,MAGA}. See also 
\cite{pisa}.

\bibitem{ALGA} J. Marro and A. Achahbar, J. Stat. Phys. {\bf 90}, 817
(1998).     

\bibitem{MAGA} J. Marro, et al.
% A. Achahbar, P. L. Garrido, and J. J. Alonso,
Phys. Rev. {\bf 53}, 6038 (1996).

\bibitem{RDDS} B. Schmittmann and R. K. P. Zia, Phys. Rev. Lett. {\bf 66},
 357 (1991).  B. Schmittmann, Europhys. Lett. 
{\bf 24}, 109 (1993). See also \cite{zia} section 6.1. 

\bibitem{RDDS2} E. Praestgaard, et al.
% H. Larsen, and R. K. P. Zia,
Europhys. Lett. {\bf 25}, 447 (1994). See also, 
E. Praestgaard, et al., cond-mat/0010053.
 
\bibitem{2T} P. L. Garrido, J. L. Lebowitz, C. Maes, and H. Spohn,
Phys. Rev. A {\bf 42}, 1954 (1990).  
                                                               
\bibitem{Binder} K. Binder, Z. Phys. B {\bf 43} 119 (1981).  

\bibitem{sf} The scaling function $f(x)$ needs to obey
$f(x) \sim x^\beta$ ($f(x) \sim x^{-(\gamma/2)}$) for large $x$
in the supercritical (subcritical) branch.

\bibitem{metric} 
%V. Privman and M. E. Fisher, Phys. Rev. B {\bf 30}, 322 (1984);
V. Privman, et al.
% P. C. Hohenberg, and A. Aharony,
  {\it Phase Transitions and Critical Phenomena}, 
Ed. C. Domb and J. L. Lebowitz (London Academic 1991), Vol. 14;
% P. C. Hohenberg  + A. Aharony 
% 
%Y. Okabe and M. Kikuchi, Int. J. Mod. Phys. C {\bf 7}, 287 (1996);
%K. Kaneda,
% Y. Okabe, and M. Kikuchi, 
%et al, J. Phys. A {\bf 32}, 7263 (1999).

\bibitem{H} M. Henkel and U. Schollw\"ock, Condmat/0010061.

\bibitem{Korea} 
S. Kwon et al. Phys. Rev. Lett. {\bf 85}, 1682 (2000).

\bibitem{pisa} S. Caracciolo, et al. Preprint 2001.
Condmat/0106221.

%In systems with absorbing states the presence of
%(apparently innocuous) parity conservation changes the universality class
%\cite{MD}. 

%\bibitem{PP}
%F. de los Santos and P.L. Garrido, {\em J. Stat. Phys.\/},
%{\bf 96}, 303 (1999).
%bibitem{triangles} See A. D. Rutenberg and C. Yeung, Phys. Rev.
%E {\bf 60}, 2710 (1999), and references therein.
%\bibitem{Geoff} G. Grinstein, J. Applied Phys. {\bf 69}, 5441 (1991).



\end{thebibliography}
\end{document}